# STEREOCHEMISTRY OF POLYPEPTIDE CONFORMATION IN COARSE GRAINED ANALYSIS

ANIL KORKUT[1#] and WAYNE A. HENDRICKSON[1,2,3]*

[1] *Department of Biochemistry and Molecular Biophysics,*
[2] *Department of Physiology and Cellular Biophysics, and*
[3] *Howard Hughes Medical Institute,*
*Columbia University, New York, NY 10032 USA*

* E-mail: wayne@xtl.cumc.columbia.edu
# Current address: Computational Biology Center, Memorial Sloan-Kettering Cancer Center, New York, 10065, USA

The conformations available to polypeptides are determined by the interatomic forces acting on the peptide units, whereby backbone torsion angles are restricted as described by the Ramachandran plot. Although typical proteins are composed predominantly from α-helices and β-sheets, they nevertheless adopt diverse tertiary structure, each folded as dictated by its unique amino-acid sequence. Despite such uniqueness, however, the functioning of many proteins involves changes between quite different conformations. The study of large-scale conformational changes, particularly in large systems, is facilitated by a coarse-grained representation such as provided by virtually bonded Cα atoms. We have developed a virtual atom molecular mechanics (VAMM) force field to describe conformational dynamics in proteins and a VAMM-based algorithm for computing conformational transition pathways. Here we describe the stereochemical analysis of proteins in this coarse-grained representation, comparing the relevant plots in coarse-grained conformational space to the corresponding Ramachandran plots, having contoured each at levels determined statistically from residues in a large database. The distributions shown for an all-α protein, two all-β proteins and one α+β protein serve to relate the coarse-grained distributions to the familiar Ramachandran plot.

## 1. Introduction

Accurate understanding of the relationship between protein structure and biological function has been the central problem in biophysics since the dawn of structural biology[1]. Macromolecular crystallography has provided thousands of unique protein structures and that have helped to explain, at an atomic level of detail, diverse biological functions such as oxygen transport[2], protein folding[3], ATP synthesis[4], ion conduction across membranes[5], protein translation[6-8], RNA transcription[9], viral entry to host cells[10], and many more. Such triumphs of structural biology also demonstrate diverse tertiary structures, each one folded into a well-defined structure determined by its unique amino-acid sequence[11]. Although the overall foldings of proteins prove to be highly varied, locally it is the polypeptide conformations of helices and sheets that predominate as organized into regular secondary structures.

The actuality of polypeptides folding into regularly repeated conformations was anticipated brilliantly by Pauling and coworkers who proposed detailed structures of the α-helix[12] and β-pleated sheets[13] based on dimensions from crystal structures of amino acids and short peptides and principles of hydrogen bonding. The spatial distribution of hydrogen bonds between backbone amide and carboxyl groups and the planar nature of the peptide bonds determined the recurrent local structural elements of helices and strands in sheets. In return, the secondary structure elements have well defined geometries with energetically preferred dihedral angles along the polypeptide main chain. The analysis of this polypeptide geometry (i.e. the $\varphi$ and $\psi$ dihedral angles of polypeptide main chain, Figure 1A) by G. N. Ramachandran and colleagues resulted in the determination of stereochemical criteria for polypeptide chains and one of the most widely used structural analysis tools in biology, the Ramachandran plot[14]. Subsequently, the stereochemistry of polypeptide conformation was placed on a firm theoretical foundation by considering the energetic consequences of the interdependent restraints from bond distances, bond angles, dihedral angles, non-bonded interactions and the spatial organization of hydrogen bonds[15].

The static description of protein structures is insufficient to explain protein function. Descriptions of conformational transitions and allosteric regulation are essential to understand the relation between protein structure and function. For this purpose, macromolecular crystallography provided snapshots of proteins in distinct conformational (and activity) states and computational biophysics provided methods such as molecular dynamics simulations. However, atomic-level analysis of large conformational transitions, particularly in large macromolecular complexes such as ribosome particles[6-8], is still computationally infeasible. Thus, a reduction in the structural complexity is required. Based on this motivation, coarse grained approaches that provide normal mode evaluation of local protein fluctuations have become popular[16]; however, in order to follow the time course of large-scale molecular events, more is required. A successful coarse-grained approach for the simulation of protein dynamics has three crucial requirements: (i) an accurate representation of polypeptides with reduced complexity, (ii) a force field that captures the molecular restraints relevant to the coarse-grained representation, and (iii) a molecular mechanics algorithm that uses the reduced representation and the potential function to accurately and efficiently compute molecular dynamics.

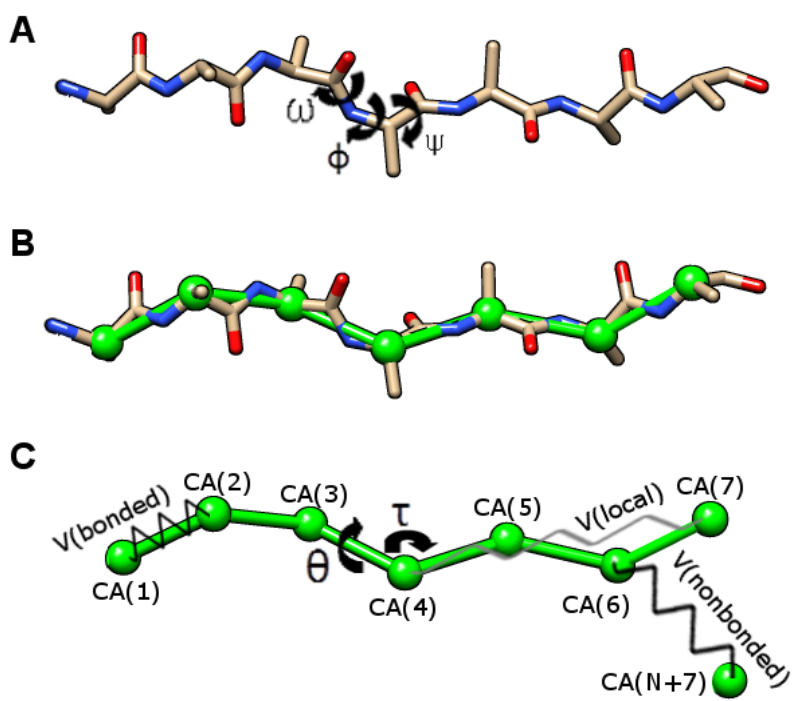

**Figure 1 Polypeptide main chain and coarse grained modeling.** A. The full-atom polypeptide main chain with dihedral angles (φ,ψ,ω). B. Conversion of the full-atom main-chain model to the Cα-atom coarse-grained representation. C. The coarse-grained representation in VAMM force field with virtual angles θ and τ.

## 2. Virtual atom molecular mechanics

For our analyses we have used the Cα virtual atom representation[17] to reduce complexity (Figure 1B). Then, to facilitate simulations, we developed virtual atom molecular mechanics (VAMM) force field[18] for coarse-grained analyses and a VAMM-based algorithm to compute conformational transition pathways for protein molecules[19]. VAMM is currently based on Cα-only representation of protein molecules similar to that in elastic network models[16] and can be extended to include side-chain centroids as in other coarse-grained models[20]. The VAMM force field restrains a given protein structure through use of virtual (i.e. Cα- Cα) bond length, angle bending, dihedral, non-bonded, and local interaction terms (Figure 1C, Equation 1). Each restraint is parameterized statistically against crystallographic data with the Boltzmann conversion method.

$$V_{VAMM} = V_{bond} + V_{angle} + V_{dihedral} + V_{nonbonded} + V_{local} \qquad (1)$$

In VAMM, we determined secondary structure specific virtual dihedral ($\theta$) and bond angle ($\tau$) restraints based on the individual distributions of every possible configuration of dihedral angle quartets (e.g. HHHH for four consecutive residues with helical structure) and bond angle triplets (e.g. EEE for three consecutive residues with β-sheet structure). The analysis of the virtual dihedral and bond angle distributions also revealed that Cα backbones of proteins are specified with ($\theta$, $\tau$) virtual angles as the main chain is defined by ($\varphi$, $\psi$) dihedrals. In analogy to the Ramachandran plot, we described probability contours of ($\theta$, $\tau$) and ($\theta$-, $\theta$+) at levels of 95% (favored) and 99% (allowed) based on the statistical information in the Top500 structure database (See methods).

Here, we further analyze the virtual ($\theta$, $\tau$) angle distributions in different proteins to demonstrate how such contour plots differ from each other for distinct folds and how they relate to the Ramachandran plot. The analysis reveals well defined regions on the ($\theta$, $\tau$) and ($\theta$-, $\theta$+) plots, which are populated by distinct secondary structure elements similar to the case on Ramachandran plot. The overall analysis demonstrates the coarse-grained stereochemical properties of a peptide chain are directly related to the stereochemical properties of the main chain.

### 3. Results

In order to determine the stereochemical properties of polypeptides in coarse-grained representation, we constructed probability contour plots of virtual angle parameters and analyzed a set of protein structures with diverse secondary structure elements (Figure 2). Analyzed structures are TorS$_S$ (all α-helices), the sensor domain of a histidine kinase receptor taken from the complex of TorS$_S$ with the ligand-binding co-receptor TorT [21], histidine kinase sensor domain HK3$_S$ [22] (all β-strands), glycoprotein CD4 [23] (D1D2 immunoglobulin-like domains, all β-strands), HIV envelope glycoprotein gp120 [23] (mixture of α-helices/β-strands). For each protein molecule, the Cα (θ, τ) and (θ-, θ+) distributions are plotted on a contour plot as in Ref. 19 and compared to the Ramachandran plot.

### *3.1. Virtual bond angles are more constrained than virtual dihedrals*

The analysis of the (θ, τ) contour plots reveals that the virtual bond angles are more constrained than the virtual dihedral angles. 99% of the measured virtual dihedral angles are confined between 75° and 148°, regardless of the secondary structure in their vicinity. On the other hand, the virtual dihedral angles sampled in protein structures span all possible angles between 0° and 360°. Thus, in the (θ, τ) distribution only a fraction of the plot is populated in the τ dimension, whereas there is a continuous distribution in the θ dimension. Nevertheless, each virtual dihedral is still constrained by the local structural preferences of its immediate vicinity. This constraint is depicted in the (θ-, θ+) contour plots such that the regions (300° < θ- < 360°, 300° < θ+ < 360°) and (75° < θ- < 150°, 75° < θ+ < 150°) are predominantly disallowed according to the distribution of virtual dihedrals sampled in Top500 database. Note that polypeptide segments involving glycine residues more frequently sample the space 75° < θ- < 150° compared to the segments without glycine (data not shown).

### *3.2. Proteins have characteristic (θ, τ) and (θ-, θ+) distributions*

The analysis of the θ and τ angles shows that each protein has a unique (θ, τ) and (θ-, θ+) distribution (Figure 2). In the all α-helical protein, TorS$_S$ (Figures 2A-D), the (θ, τ) plot is mainly populated around the peak (θ = 50°, τ = 90°) and few points are located outside this peak. Similarly, the peak around (θ- = 50°, θ+ = 50°) singles out on the (θ-, θ+) plot. Not surprisingly, the region corresponding to the α-helix on Ramachandran plot is most populated for TorS$_S$. The all β-strand protein CD4 (Figures 2I-L) shows a remarkably different profile with a major peak centered around (θ = 195°, τ = 117°) and a significant number of points extending outside this region to lower dihedral angle values. The structural analysis shows that these points correspond to those residues residing on the loops. The (θ-, θ+) distribution has a sharp peak around (θ- = 195°, θ+=195°). The significantly larger HK3$_S$ structure (all β-strand, 739 residues,

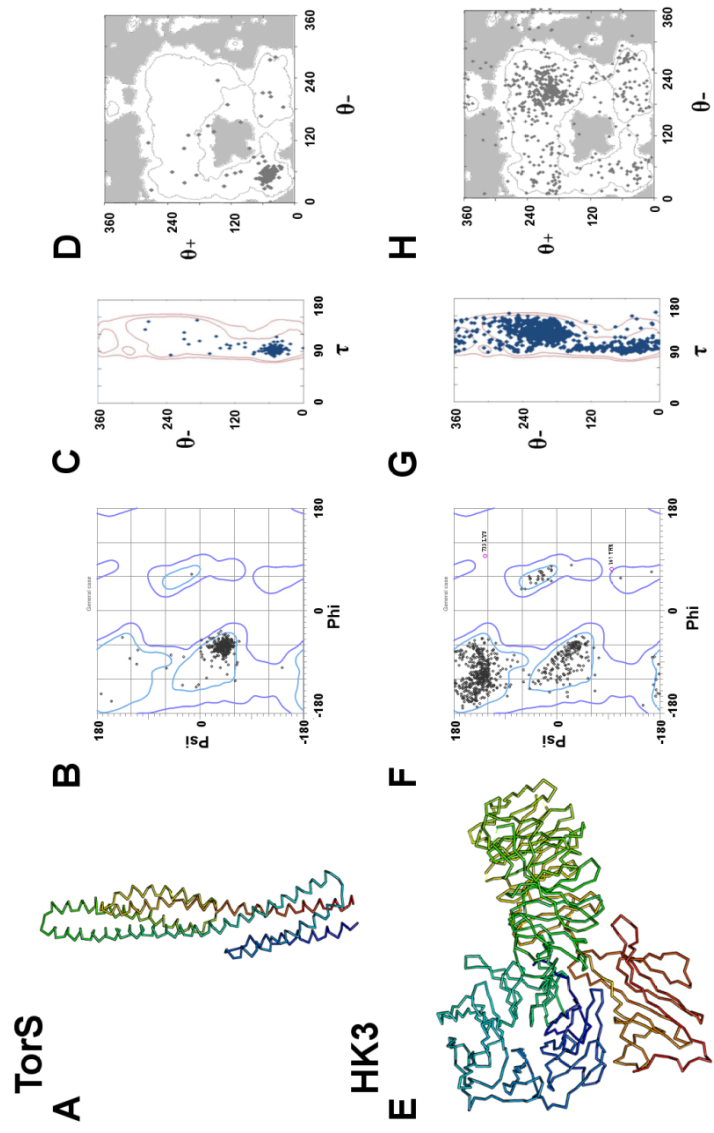

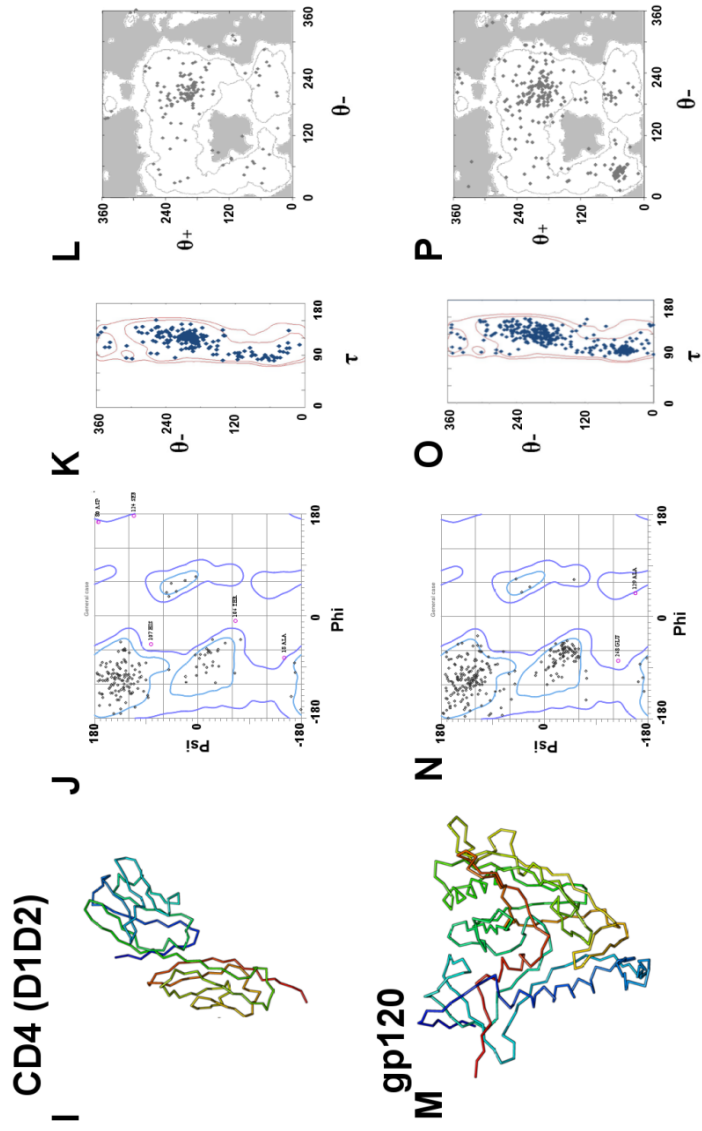

**Figure 2. Stereochemical analysis of protein structures.** A. The structure of TorS$_S$. B. Ramachandran plot for TorS$_S$. C. ($\theta$, $\tau$) plot for TorS$_S$. D. ($\theta-$, $\theta+$) plot for TorS$_S$. E. The structure of HK3. F. Ramachandran plot for HK3$_S$. G. ($\theta$, $\tau$) plot for HK3$_S$. H. ($\theta-$, $\theta+$) plot for HK3$_S$. I. The structure of CD4-D1D2. J. Ramachandran plot for CD4. K. ($\theta$, $\tau$) plot for CD4. L. ($\theta-$, $\theta+$) plot for CD4. M. The structure of HIV gp120. N. Ramachandran plot for gp120. O. ($\theta$, $\tau$) plot for gp120. P. ($\theta-$, $\theta+$) plot for gp120.

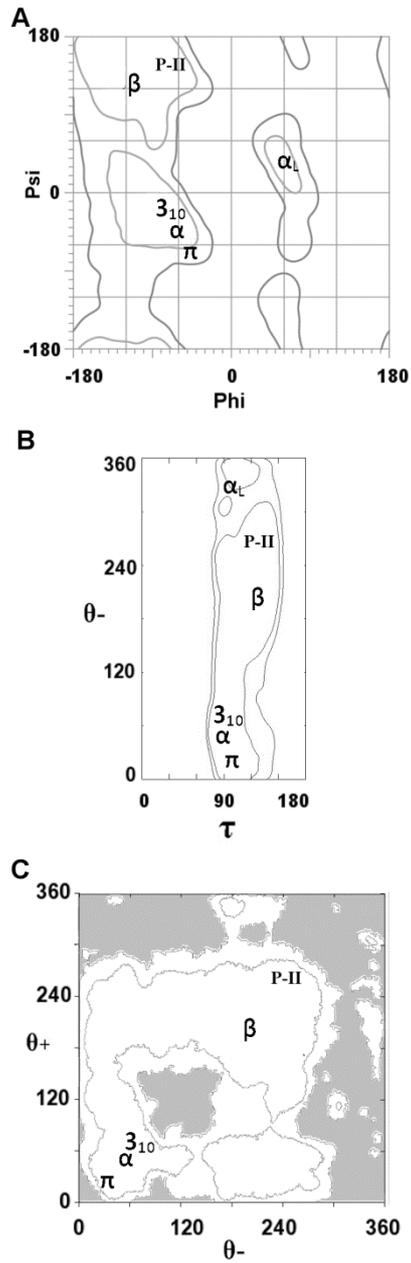

**Figure 3. Distribution of secondary structure elements.** Each secondary structure element (β-strand, α-helix, $α_L$, $3_{10}$ helix, and π-helix) populates a specific region in actual and virtual dihedral and bond angle spaces. The position of each element is shown on each plot: A. The Ramachandran plot. B. The (θ, τ) plot. C. The (θ-, θ+) plot. Note that the (θ, τ) distributions of peptides in $α_L$ configurations were computed using short polypeptide segments (4 residues) extracted from the PDB as given by Novotny & Kleywegt[25]. There are no long $α_L$ segments, and short fragments do not allow for computation of (θ-, θ+).

Figures 2E-H) shares the same peaks with CD4. Most interestingly, a large number of points are observed outside this region extending to the helical zone on both (θ, τ) and (θ-, θ+). The same characteristics are also observed in the Ramachandran plot (Figure 2F). These points correspond to residues on the loops connecting the strands of the β-propeller domains, which retain short helical or turn structures (e.g. a $3_{10}$ helix at Pro776-His781 and an α-helix at Asp165-Asn168 according to a DSSP analysis). Not surprisingly, the (θ, τ) and (θ-, θ+) plots of the α/β mixed gp120 displays both of the peaks observed in all helical and β-sheet proteins.

### 3.3. Each secondary structure element populates a distinct region on θ/τ plots

The unique θ/τ distributions for each protein and the VAMM force field parameterization[18] have suggested that we can define stereochemical criteria for coarse grained analysis of polypeptide chains. Such criteria are analogous to those seen in Ramachandran plots (Figure 3A), where positions of regularly repeated secondary structure are now refined from those identified initially[14] through careful analyses of the PDB[24-26] (Table 1). A detailed analysis of θ/τ distributions demonstrates how each secondary structure element occupies a distinct area on contour plots (θ, τ) and (θ-, θ+) (Table 1, Figures 3B-C).

Table 1. Approximate preferred virtual angles for secondary structure elements

| Secondary Structure | (φ, ψ) | (θ, τ) | (θ-, θ+) |
|---|---|---|---|
| α-helix | (-63°, -43°)[24] | (50°, 90°) | (50°, 50°) |
| β-strand | (-116°, 129°)[24] | (194°, 120°) | (194°,194°) |
| Polypeptide-II | (-65°, 145°)[24] | (249°, 118°) | (249°, 251°) |
| $3_{10}$-helix | (-62°, -22°)[24] | (67°, 90°) | (67°, 67°) |
| $α_L$-helix | (59°, 42°)[25] | (320°, 90°) | (X, 320°) |
| π-helix | (-76°, -41°)[26] | (22°,103°) | (22°, 22°) |

### 4. Discussion

Here we have analyzed the stereochemistry of polypeptides for coarse grained analysis. The distribution of virtual dihedral and bond angles in Cα-traces shows that there are a set of stereochemical criteria that distinguishes each secondary structure type in coarse grained representation. In analogy to the Ramachandran plot, (θ, τ) and (θ-, θ+) probability contour plots define favored and allowed regions for polypeptides. These contour plots can serve to determine the quality of structures in their energy-minimum conformations and transition states computed with coarse-grained simulations. Such plots will be useful in analysis of conformation transition pathways and guide us to develop more accurate coarse grained modeling strategies.

## 5. Methods

### 5.1. Protein structures

All protein structures were obtained from the Protein Data Bank (www.pdb.org). The TorS$_S$ chain A is stripped from the TorS$_S$-TorT complex structure at 2.80 Å resolution (PDB id: 3O1I). The HK3$_S$ structure is chain A of the dimeric structure of the extracellular domain of putative one component system BT4673 at 2.3 Å resolution (PDB id: 3OTT). The gp120 and CD4(D1D2) structures are extracted from the structure of the gp120 core complexed with the membrane-distal immunoglobin-like (Ig) domains D1 and D2 of CD4 and an Fab fragment of antibody 17b at 2.2 Å resolution (PDB id: 1RZJ).

### 5.2. Construction of the (θ, τ) and (θ-, θ+) plots

The favored and allowed regions of the virtual dihedral (θ) and bond angles (τ) were computed by a procedure similar to that used in MolProbity software to compute the contours of the Ramachandran plot[27]. The contour levels were determined statistically against the Top500 database, which was also used to construct Ramachandran plots in MolProbity. Different from the Ramachandran plots, the Gly and Pro residues were not analyzed separately. The Top500 database contains structures determined at better than 1.8 Å resolution and residues with B-factors higher than 40 are not included in the statistical analysis. A density dependent smoothening function[28] was used to generate contour plots such that the sparse regions of the distributions are represented smoothly and continuous, and the sharp transitions are preserved. All of the scripts and templates for generating the (θ, τ) and (θ-, θ+) plots are available at http://www.virtualatom.org.

**Acknowledgement**

This work was supported in part by National Institutes of Health Grant GM56550 (to W. A. H.).